\documentclass[ reprint,aps]{revtex4-2}
\usepackage{amsfonts}
\usepackage{eurosym}
\usepackage{amssymb}
\usepackage{amsmath}
\usepackage{tabularx,epsfig,color,tabularx,epstopdf}
\usepackage{graphicx}
\usepackage{dcolumn}
\usepackage{bm}

\begin{document}

\title{Dissipative structures in one- and two-dimensional Kerr cavities with
a spatially periodic pump}
\author{Wen-Rong Sun$^{1}$}
\email{sunwenrong@ustb.edu.cn}
\author{Wei-Chao Ma$^{1}$}
\author{Boris A. Malomed$^{2,3}$}
\affiliation{$^{1}$ School of Mathematics and Physics, University of Science
and Technology Beijing, Beijing 100083, China\\
$^{2}$ Department of
Physical Electronics, School of Electrical Engineering, Tel Aviv University,
Tel Aviv 69978, Israel\\
$^{3}$Instituto de Alta Investigaci\'{o}n,
Universidad de Tarapac\'{a}, Casilla 7D, Arica, Chile}

\begin{abstract}
The interplay of periodic driving and dissipation is a fundamental feature
of nonequilibrium physics. We elaborate a scenario for the formation of
dissipative multi-spot excitations (MSEs) in Kerr cavities, modeled by the
one- and two-dimensional (1D and 2D) Lugiato-Lefever (LL) equations, which
include a spatially periodic pump (SPP). First, we demonstrate that the SPP
produces three novel exact periodic solutions of the LL equation, expressed
in terms of the $\mathrm{sn}$, $\mathrm{cn}$, and $\mathrm{dn}$ elliptic
functions. By means of numerical methods, we explore the modulational
instability (MI) and transverse instability (TI) of the periodic states in
1D and 2D settings, respectively. In the case of the defocusing
nonlinearity, the 1D MI breaks the periodic states into an array of
spatiotemporal crescents. In the case of self-focusing, the 1D MI, initiated
by small random perturbations, leads to establishment of a chaotic state,
with the amplitude statistics featuring a long-tail probability
distribution, that represents the presence of dissipative rogue waves. On
the other hand, spatially periodic perturbations initiate formation of
breather chains, which periodically disappear and reappear, resembling the
Fermi-Pasta-Ulam-Tsingou recurrence. In the 2D regime, the TI results in the
formation of an array of 2D lumps. For a given SPP strength, the exact
solutions for the periodic structures are stable if the loss constant
exceeds a critical value. The findings reported here provide a contribution
to nonequilibrium physics in general and may find\ direct applications in
laser physics.
\end{abstract}

\maketitle








\section{Introduction}

Multi-spot excitations (MSEs) are far-from-equilibrium structures that
dominate dynamics in various settings in physics, chemistry, biology,
\textit{etc}.~\cite{ll1,ll2,Aranson,xinjia1}. Nonlinear optics and laser
physics provide basic examples of MSEs that have been subjects of intensive
theoretical and experimental studies \cite%
{nolp1,nolp2,nolp3,nolp4,nolp5,nolp6,nolp7,nolp8,Purwins,Staliunas}. In
dissipative systems, the formation of MSEs, such as those built of
dissipative solitons, is underlain by the balance between the intrinsic loss
and externally applied gain or pump~\cite{PhysicaD,nolp9,nolp10}. Based on
complex Ginzburg-Landau equations, theoretical and experimental studies of
dissipative solitons and MSEs have been conducted in various contexts~\cite%
{nolp11,nolp8}. In nonlinear optical systems including an external drive,
the losses are compensated by the pump supplied by illuminating laser beams.
The Lugiato-Lefever (LL) equation is the fundamental model of such driven
systems~\cite{ll3}. One- and two-dimensional (1D and 2D) LL equations have
been used to predict various states~\cite{ll4,ll5,ll6,ll7,ll8}, such as Kerr
solitons, frequency combs, breathers and rogue waves (RWs)~\cite%
{an1,an2,an3,an4,an5,an6,an7,an8,an9,an10,Science,Tlidi,an11,an12,an13,an17,an18,an19,an20}%
. Ref.~\cite{mt1} presented the evidence of 1D and 2D MSEs for the LL
equation, which describes nonlinear optical resonators subjected to a
continuous laser pump. Also relevant in the context of the generation of
MSEs are Refs. \cite{mt2} and \cite{mt3}. Experimental evidence of soliton
formation resulting from the front interaction, modeled by the LL equation
including an inhomogeneous pump laser, was presented in Ref.~\cite{mt4}.

This work aims to reveal a new scenario for the formation of dissipative
MSEs in Kerr cavities by considering the interplay of a spatially periodic
pump (SPP) and dissipation, which is a fundamental setting in nonequilibrium
physics~\cite{ll1,ll2,xinjia1}. First, we demonstrate \ that SPP excites
novel exact periodic solutions of the 1D LL equation, which are expressed in
terms of elliptic functions \textrm{$sn$}, \textrm{$cn$}, and \textrm{$dn$}.
Then, using the Floquet-Fourier-Hill theory~\cite{ffh1}, we explore the
modulational instability (MI) of the periodic solutions in the 1D case. The
result, produced by the numerical solution of the respective linear spectral
problem, is asymmetric with respect to the imaginary axis in the plane of
the stability eigenvalues. In the case of the defocusing nonlinearity, the
asymmetric MI breaks the underlying \textrm{$sn$ }pattern\textrm{\ }into an
array of spatiotemporal crescents. In the case of self-focusing, the
asymmetric MI of the \textrm{$cn$} and \textrm{$dn$ }patterns leads to the
formation of dissipative RWs. In this case, the amplitude probability
distribution function features a long tail, which implies that the local
intensity in the RWs exceeds twice the significant wave height (SWH, alias
the background level)~\cite{rwww2}. We also demonstrate recurring formation
of chains of breathers, as a manifestation of the Fermi-Pasta-Ulam-Tsingou
(FPUT) recurrence. In the 2D regime, the transverse instability (TI) of the
quasi-1D periodic patterns occurs, leading to generation of arrays of 2D
lumps. We find critical (threshold) values of the strength of the periodic
pump, $E_{(0,\mathrm{sn/cn/dn})}$, below which the respective stationary
patterns are stable, for a given value of the loss constant. This setting
and MSEs supported by it have not been previously reported. The present
results improve the understanding of the interplay of periodically
structured pumps and dissipation in the general case.

The subsequent presentation is structured as follows. The basic LL equation,
its exact spatially-periodic analytical solutions, on which the work is
focused, and the spectral problem for the study of their stability, are
formulated in Section 2. The basic results for the 1D states, including
their stability and evolution of unstable ones, are reported in Sections 3
and 4, for the repulsive and attractive signs of the cubic nonlinearity,
respectively. The results for the stability analysis and evolution of
unstable states in the 2D LL equation with the quasi-1D SPP are briefly
summarized in Section 5. The paper is concluded by Section 6.

\section{The model, exact solutions and spectral problem}

The LL equation for amplitude $\phi (x,t)$ of the electromagnetic field in a
nonlinear lossy cavity driven by SPP $E(x)$ is written as
\begin{equation}
i\left( \gamma +\frac{\partial }{\partial t}\right) \phi =\left[ -\frac{1}{2}%
\frac{\partial ^{2}}{\partial x^{2}}+\Delta +\sigma |\phi |^{2}\right] \phi
+E(x),  \label{lle}
\end{equation}%
where $\gamma >0$ is the loss constant, $\Delta \gtrless 0$ is detuning of
the pump with respect to the cavity, while $\sigma =+1$ and $\sigma =-1$
corresponds to the defocusing and focusing nonlinearity, respectively. It is
relevant to mention that the laser cavity modeled by Eq. (\ref{lle}) can be
pumped by two drives, \textit{viz}., the external one, represented by $E(x)$%
, and internal, which creates intrinsic gain $\Gamma $ \cite{gain1,gain2},
so that original $\gamma $ is replaced by the residual loss parameter, $%
\gamma _{\mathrm{residual}}=\gamma -\Gamma $. This circumstance makes it
possible to vary the effective loss in the experiment.

To understand the interplay of the SPP, loss, paraxial diffraction,
accounted for by the second derivative in Eq.~(\ref{lle}), and the
nonlinearity, stationary analytical solutions of Eq.~(\ref{lle}) are needed.
To this end, we set $E(x)=E_{(0,\mathrm{sn})}$\textrm{$sn$}$(x,k)$ in the
defocusing regime, and $E(x)=E_{(0,\mathrm{dn})}$\textrm{$dn$}$(x,k)$ or $%
E_{(0,\mathrm{cn})}$\textrm{$cn$}$(x,k)$ in the focusing one, where \textrm{$%
sn$} and \textrm{$dn$ }or\textrm{\ $cn$} are the Jacobi's elliptic functions
with modulus $k$, and $E_{(0,\mathrm{sn/dn/cn})}$ is the strengths of the
respective pump. These patterns are determined by the classical Fourier
representation \cite{hb1,Gradstein}. For instance, it is%
\begin{equation}
{\mathrm sn}(x;k)=\frac{2\pi }{kK\left( k\right) }\sum_{n=0}^{\infty }\frac{%
q^{n+1/2}}{1-q^{2n+1}}\sin \frac{(2n+1)\pi \alpha x}{2K\left( k\right) },
\label{sn}
\end{equation}%
where $q=\exp \left\{ -\pi \left[ K\left( \sqrt{1-k^{2}}\right) /K\left(
k\right) \right] \right\} $ and $K(k)$ is the complete elliptic integral of
the first kind. In fact, for $k<0.9$ the periodic pumps, such as the one
represented by expression (\ref{sn}), can be well approximated by only a few
spatial harmonics~\cite{xiugai1,Carr1}, which facilitates the creation of
such profiles in the experiment. On the other hand, the same Jacobi's
functions represent eigenmodes of nonlinear media with the cubic
nonlinearity \cite{hb1,Gradstein}, therefore the corresponding profiles can
be produced by passing the pump beam through a layer of an appropriate
optical material.

Then, the following novel spatially periodic solutions of~Eq. (\ref{lle})
are obtained:

$\bullet $ In the defocusing case ($\sigma =1$ and $E(x)=E_{(0,\mathrm{sn})}$%
\textrm{$sn$}$(x,k)$),
\begin{equation}
\phi _{\mathrm{sn}}(x)=e^{i\theta _{\mathrm{sn}}}k~\mathrm{sn}(x,k),\theta _{%
\mathrm{sn}}=\tan ^{-1}\left( \frac{2\gamma }{2\Delta +k^{2}+1}\right) ,
\label{sns}
\end{equation}%
with the loss constant and SPP strength satisfying the constraint
\begin{gather}
E^2_{(0,\mathrm{sn})}=\frac{1}{4}k^2 ({4\gamma ^{2}+(2\Delta
+1)^{2}+(4\Delta +2)k^{2}+k^{4}}).  \label{dyc}
\end{gather}

$\bullet $ In the focusing case [$\sigma =-1$ and $E(x)=E_{(0,\mathrm{cn})}$%
\textrm{$cn$}$(x,k)$ or $E_{(0,\mathrm{dn})}$\textrm{$dn$}$(x,k)$],
\begin{equation}
\phi _{\mathrm{cn}}(x)=e^{i\theta _{\mathrm{cn}}}k~\mathrm{cn}(x,k),\theta _{%
\mathrm{cn}}=\tan ^{-1}\left( \frac{2\gamma }{2\Delta -2k^{2}+1}\right) ;
\label{cns}
\end{equation}%
\begin{equation}
\phi _{\mathrm{dn}}(x)=e^{i\theta _{\mathrm{dn}}}\mathrm{dn}(x,k),\theta _{%
\mathrm{dn}}=\tan ^{-1}\left( \frac{2\gamma }{2\Delta +k^{2}-2}\right) ,
\label{dns}
\end{equation}%
with the respective constraints%
\begin{equation}
E_{(0,\mathrm{cn})}^{2}=\frac{1}{4}k^{2}(4\gamma ^{2}+(2\Delta
+1)^{2}-4(2\Delta +1)k^{2}+4k^{4});
\end{equation}%
\begin{equation}
E_{(0,\mathrm{dn})}^{2}=\frac{1}{4}(4\left( \gamma ^{2}+(\Delta
-1)^{2}\right) +4(\Delta -1)k^{2}+k^{4}).\label{swradd}
\end{equation}%
These solutions are periodic in $x$, with period $L=4K(k)$ for solutions~(%
\ref{sns}) and~(\ref{cns}), and $L=2K(k)$ for (\ref{dns}), where $K(k)$ is
the complete elliptic integral of the first kind.

To examine the (in)stability of these stationary states, perturbed solutions
of Eq.~(\ref{lle}) are looked for as
\begin{equation}
\phi (x,t)=e^{i\theta }\psi (x)+\epsilon \left[ (u(x)e^{\lambda t}+\mathrm{%
c.c.})+i(v(x)e^{\lambda t}+\mathrm{c.c.})\right] ,  \label{lsp0}
\end{equation}%
where $\psi (x)=k~\mathrm{sn}(x,k)$, $k~\mathrm{cn}(x,k)$ and $\mathrm{dn}%
(x,k)$ respectively, $\theta =\theta _{\mathrm{sn}}$, $\theta _{\mathrm{cn}}$
and $\theta _{\mathrm{dn}}$ respectively, $\epsilon $ is a small
perturbation amplitude, $\mathrm{c.c.}$ stands for the complex conjugate,
and eigenfunctions $\left( u(x),v(x)\right) ^{T}$, corresponding to
eigenvalue $\lambda $ of the linear spectral problem, satisfy the linearized
equation
\begin{equation}
\lambda \left[
\begin{array}{l}
u(x) \\
v(x)%
\end{array}%
\right] =\left[
\begin{array}{rr}
\mathcal{L}_{11} & \mathcal{L}_{12} \\
\mathcal{L}_{21} & \mathcal{L}_{22}%
\end{array}%
\right] \left[
\begin{array}{l}
u(x) \\
v(x)%
\end{array}%
\right] .  \label{lsp1}
\end{equation}%
Here, the matrix elements are $\mathcal{L}_{11}=-\gamma +\sigma \sin
(2\theta )\psi ^{2}$, $\mathcal{L}_{12}=\Delta -\sigma \cos (2\theta )\psi
^{2}+2\sigma \psi ^{2}-(1/2)\partial _{x}^{2}$, $\mathcal{L}_{21}=-\Delta
-\sigma \cos (2\theta )\psi ^{2}-2\sigma \psi ^{2}+1(1/2)\partial _{x}^{2}$,
and $\mathcal{L}_{22}=-\gamma -\sigma \sin (2\theta )\psi ^{2}$. As $\psi (x)
$ has period $L$, the Floquet theory suggests one to look for solutions
to~Eq. (\ref{lsp1}) as
\begin{eqnarray}
u(x) &=&e^{i\mu x}U(x)=e^{i\mu x}\sum_{j=-\infty }^{+\infty }\hat{U}%
_{j}e^{2i\pi jx/PL}, \\
v(x) &=&e^{i\mu x}V(x)=e^{i\mu x}\sum_{j=-\infty }^{+\infty }\hat{V}%
_{j}e^{2i\pi jx/PL},
\end{eqnarray}%
where $U(x+PL)=U(x)$, $V(x+PL)=U(x)$, $\hat{U}_{j}=\left( PL\right)
^{-1}\int_{-PL/2}^{+PL/2}U(x)e^{-i2\pi jx/PL}dx$, $\hat{V}_{j}=\left(
PL\right) ^{-1}\int_{-PL/2}^{+PL/2}V(x)e^{-i2\pi jx/PL}dx$, and $\mu \in
\lbrack 0,2\pi /L)$ is the corresponding Floquet exponent. Here we introduce
integer $P$ to identify possible subharmonic instability (MI) of the
periodic patterns against perturbations whose spatial period, $PL$, is a
multiple of the underlying pattern's period $L$. We have found the stability
spectrum of the linear spectral problem~(\ref{lsp1}) by dint of the
Floquet-Fourier-Hill theory, the instability occurring if there is an
eigenvalue with $\mathrm{Re}(\lambda )>0$. The results are reported below.

\begin{figure}[tbp]
\centering
\includegraphics[height=90pt,width=100pt]{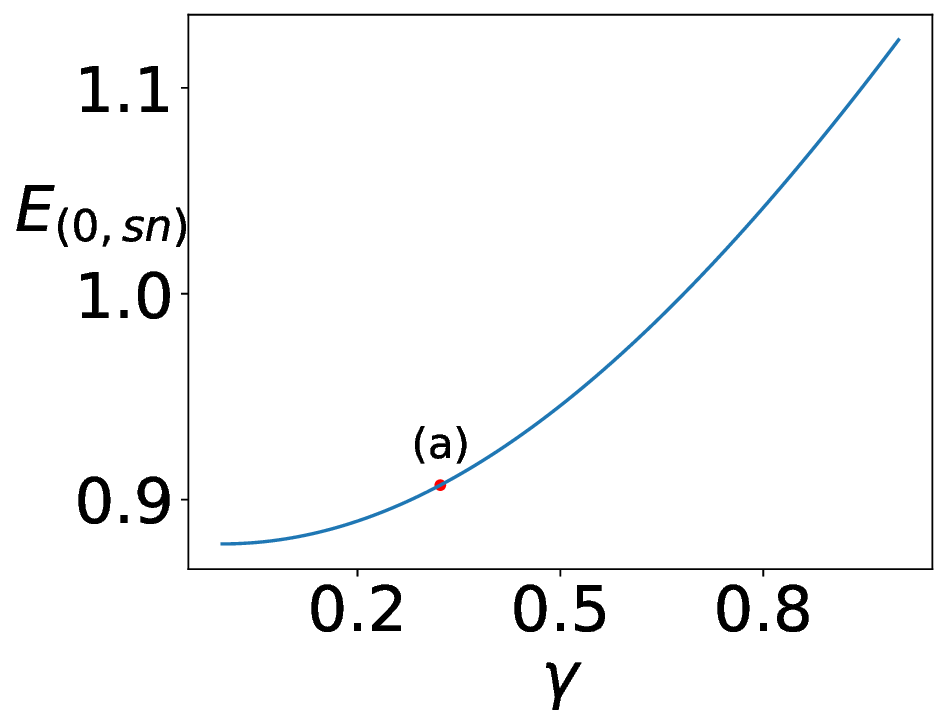} \newline
\caption{The SPP strength $E_{(0,\mathrm{sn})}$, which supports the exact
\textquotedblleft snoidal" solution (\protect\ref{sns}) as per Eq.~(\protect
\ref{dyc}), vs. the loss constant $\protect\gamma $. Here $\protect\sigma =1$%
, $\Delta =-2$ and $k=0.7$. The transition between the stable and unstable
states takes place, with the increase of $\protect\gamma $, at point $(%
\mathrm{a})$, with $\protect\gamma =0.3225$, $E_{(0,\mathrm{sn})}=0.907042)$.
}
\label{fig1}
\end{figure}

\begin{figure}[tbp]
\centering
\includegraphics[height=90pt,width=100pt]{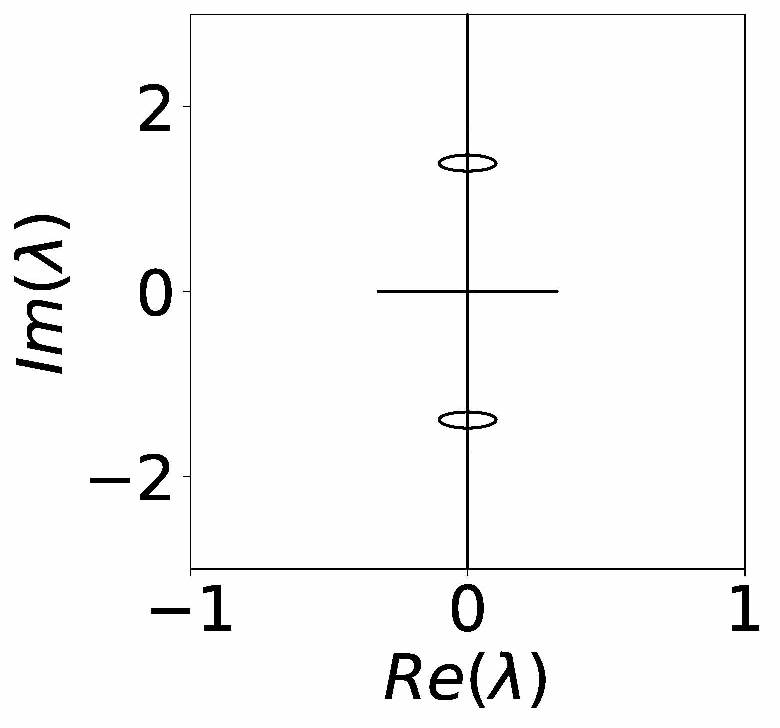} \newline
\caption{MI (modulational instability) of the \textquotedblleft snoidal
"state (\protect\ref{sns}), which is symmetric with respect to the imaginary
axis in the eigenvalue plane, in the case of $\protect\gamma =0$ and $%
\protect\sigma =1$, $\Delta =-2$, $k=0.7$.}
\label{fig2}
\end{figure}

\section{Asymmetric MI and formation of MSEs (multi-spot excitations) in the
1D defocusing regime}

In this regime, Fig.~\ref{fig1} displays relation (\ref{dyc}) between the
SPP strength ($E_{(0,\mathrm{sn})}$) and loss constant $\gamma $. The reason
we choose $\gamma $ as the control parameters is that it provides a
possibility to suppress or enhance MI~\cite{smi}, which, in turn, enables us
to control the emergence or disappearance of instabilities in the system. On
the other hand, an obvious scaling transformation of Eq. (\ref{lle}) makes
it also possible to rescale all the results into those corresponding to $%
\gamma =\mathrm{const}$ in Eq. (\ref{lle}).

\begin{figure}[tbp]
\centering
\includegraphics[height=80pt,width=80pt]{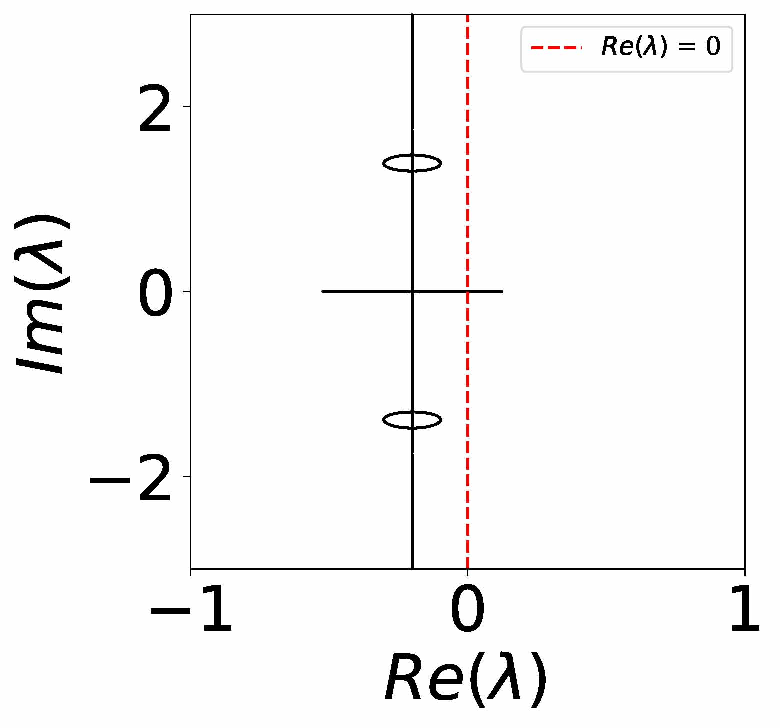} %
\includegraphics[height=80pt,width=80pt]{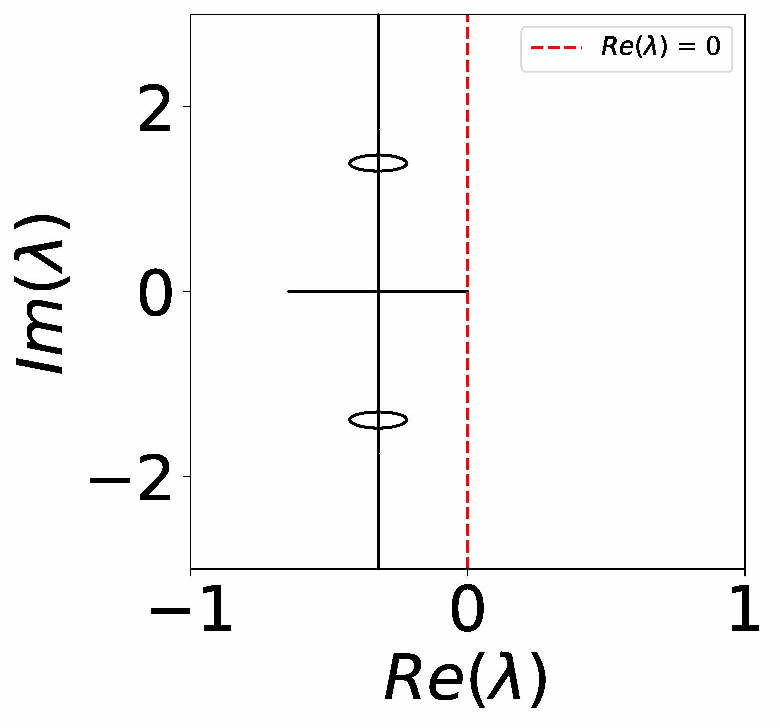} %
\includegraphics[height=80pt,width=80pt]{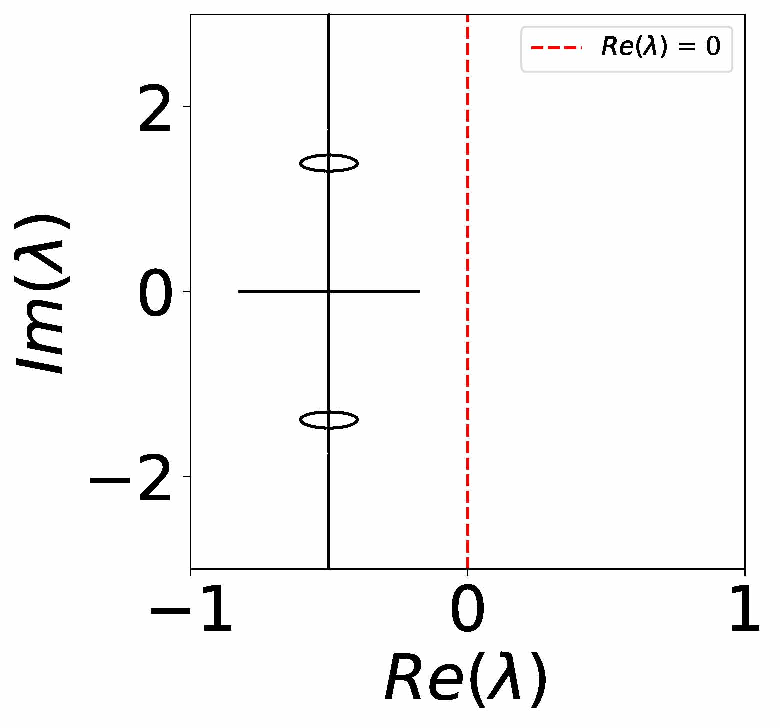} \newline
\caption{The asymmetric MI shown by the eigenvalue spectrum for the
stationary solution~(\protect\ref{sns}) with $\protect\sigma =1$, $\Delta =-2
$, $k=0.7$, and $\protect\gamma =0.2$ (the left panel: instability), or $%
\protect\gamma =\protect\gamma _{c}\approx 0.3225$ (the middle panel: the
transition from the instability to stability), or $\protect\gamma =0.5$ (the
right panel: stability).}
\label{fig3}
\end{figure}

At $\gamma =0$, the stability spectrum is symmetric with respect to the real
and imaginary axes of the plane of $\lambda $, as shown in Fig.~\ref{fig2}.
In such a case, symmetric MI arises, represented by instability bands that
include a finite segment on the real axis and two elliptical loci in the
complex plane. When the losses appear ($\gamma >0$), the spectrum as a whole
shifts to the left, breaking its symmetry with respect to the imaginary
axis, as shown in Fig.~\ref{fig3}. We find that, at $\gamma >\gamma
_{c}\approx 0.3225$ [which corresponds to point (a) in Fig.~\ref{fig1})],
all eigenvalues have Re$(\lambda )<0$, which implies the stabilization of
the \textquotedblleft snoidal" periodic pattern (\ref{sns}), see Fig.~\ref%
{fig3}. Thus, MI, which is asymmetric with respect to the imaginary axis in
the plane of $\lambda $, occurs in the defocusing system in the interval of $%
0<\gamma <\gamma _{c}$.

\begin{figure}[tbp]
\centering
\includegraphics[height=90pt,width=100pt]{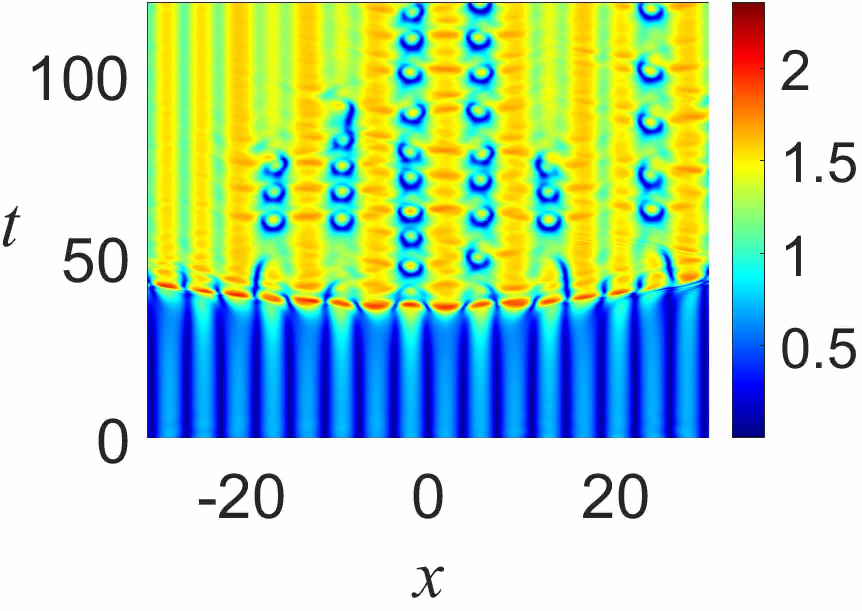} %
\includegraphics[height=90pt,width=100pt]{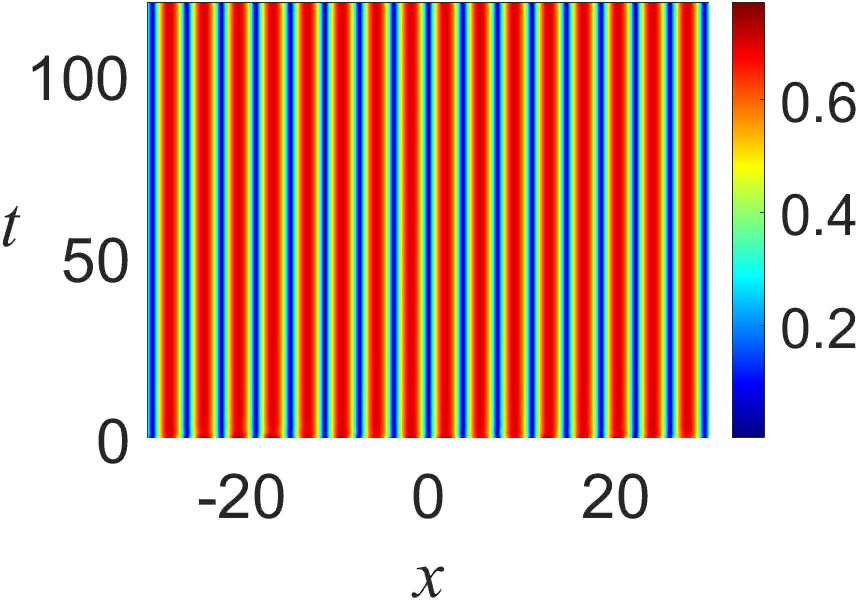} \newline
\caption{Numerical simulations demonstrating the evolution of $|\protect\phi %
(x,t)|$ for the unstable snoidal solution~(\protect\ref{sns}), initially
perturbed by $1\%$ random noise and eventually developing into an array of
small-size spatiotemporal crescents. The parameters are: $\protect\sigma =1$%
, $\Delta =-2$, $k=0.7$, and $\protect\gamma =0.2$ or $0.5$ in the left
(asymmetric MI) or right (stability) panels, respectively.}
\label{fig4}
\end{figure}

To study the development of the MI in the nonlinear regime, we performed
simulations of Eq. (\ref{lle}) by means of the split-step Fourier algorithm,
adding random-noise perturbations at the level of $1\%$ to the stationary
state (\ref{sns}). We have thus found that MI breaks the unperturbed
spatially periodic state into an array of spatiotemporal crescents, as shown
in the left panel of Fig.~\ref{fig4}. For comparison, the right panel shows
that, at $\gamma =0.5>\gamma _{c}$, the snoidal spatially periodic state
remains stable.

\begin{figure}[tbp]
\centering
\includegraphics[height=90pt,width=100pt]{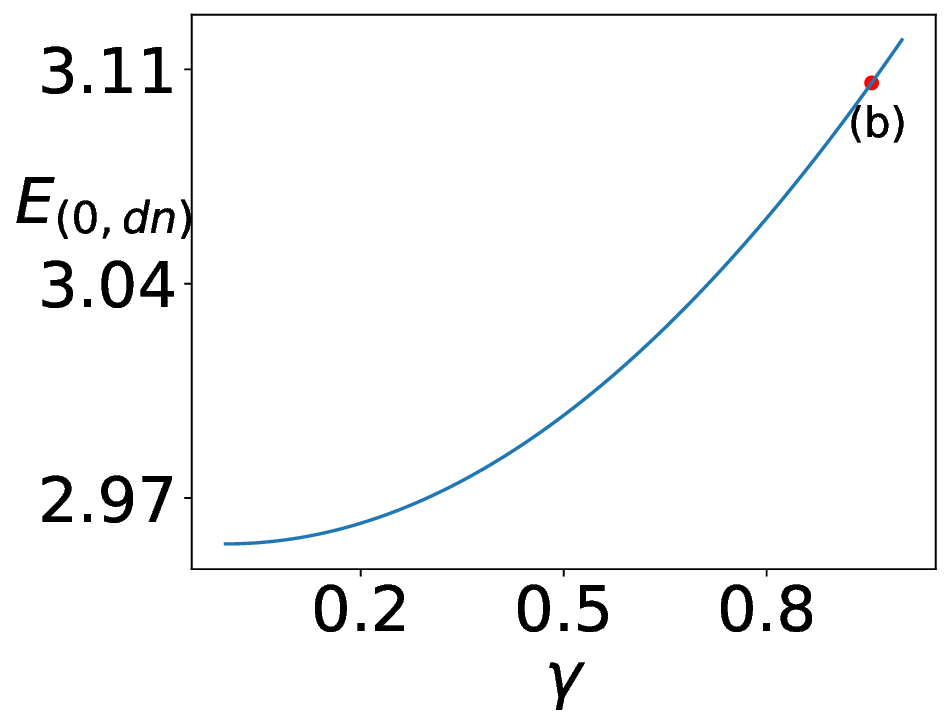} \newline
\caption{The SPP\ strength $E_{(0,\mathrm{dn})}$, which supports the exact
\textquotedblleft\ dnoidal" solution (\protect\ref{dns}) as per Eq.~(\protect
\ref{swradd}), vs. the loss constant, $\protect\gamma $. Here $\protect\sigma %
=-1$, $\Delta =-2$ and $k=0.3$. The transition between\ the stable and
unstable states takes place, with the increase of $\protect\gamma $, at
point $(\mathrm{b})$, with $\protect\gamma =0.9553$, $E_{(0,\mathrm{dn}%
)}=3.10558)$.}
\label{swbo}
\end{figure}

\section{The formation of dissipative breathers and RWs in the 1D focusing
regime. }

\begin{figure}[tbp]
\centering
\includegraphics[height=90pt,width=100pt]{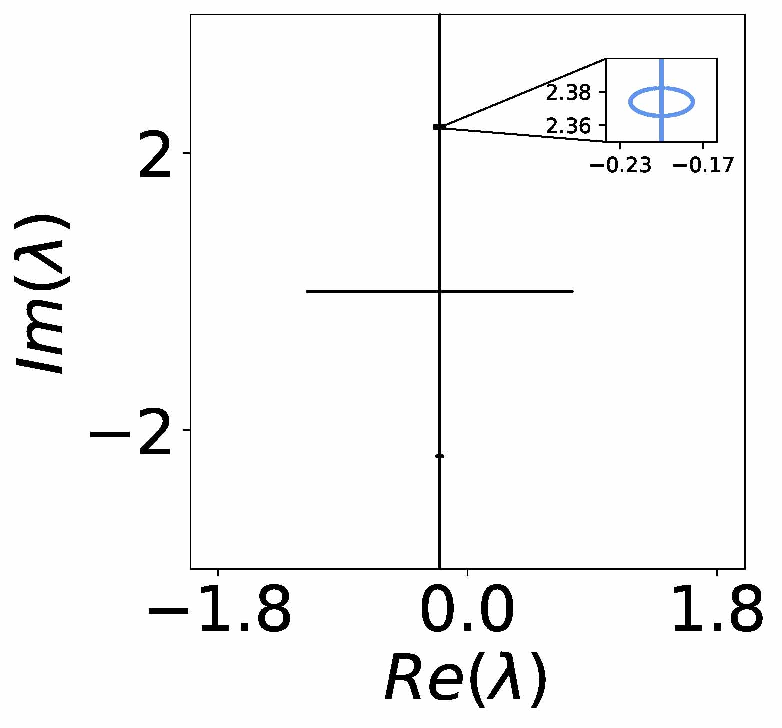} %
\includegraphics[height=90pt,width=100pt]{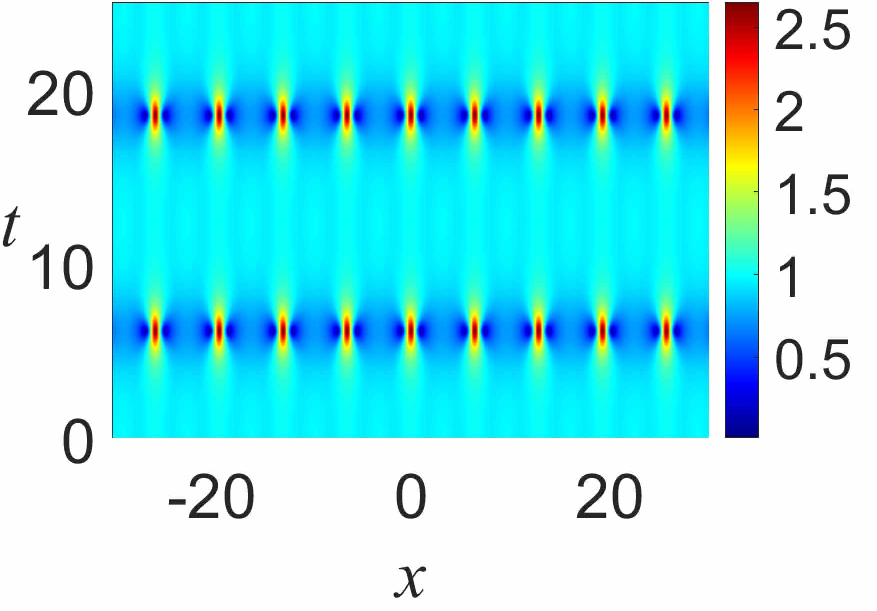} \newline
\caption{The MI, breather chains, and its FPUT-like recurrence in the case
of the focusing nonlinearity ($\protect\sigma =-1$). The left panel: the\
MI spectrum for solutions (\protect\ref{dns}) with $\Delta =-2$, $k=0.3$ and
$\protect\gamma =0.2$. The right panel: the breather chain and FPUT
recurrence, exhibited by $|\protect\phi (x,t)|$, as produced by the
simulations with the double-period perturbation added to the stationary
pattern.}
\label{fig5}
\end{figure}

Similar to the defocusing case, asymmetric MI occurs in the focusing regime,
as seen in the left panel of Fig.~\ref{fig5}, and the spatially periodic
\textquotedblleft cnoidal" and \textquotedblleft dnoidal" stationary
solutions~(\ref{cns}) and~(\ref{dns}) become stable when $\gamma $ exceeds a
critical value, as shown in Fig.~\ref{swbo}. Therefore, we do not discuss
this point in further detail here.

In the focusing regime, simulations of the MI development of the asymmetric
MI leads to generation of breathers and RWs. This dynamics being similar for
the stationary solutions~(\ref{cns}) and~(\ref{dns}), we here report
detailed results for the dnoidal state~(\ref{dns}), adding a small periodic
perturbation to it, with period $4K(k)$, which is twice that of the
unperturbed state. It is observed that a chain of breathers emerges close to
$t=7$, which subsequently disappears and reappears close to $t=18$,
resembling the FPUT recurrence, as seen in the right panel of Fig.~\ref{fig5}%
.

\begin{figure}[tbp]
\centering
\includegraphics[height=95pt,width=105pt]{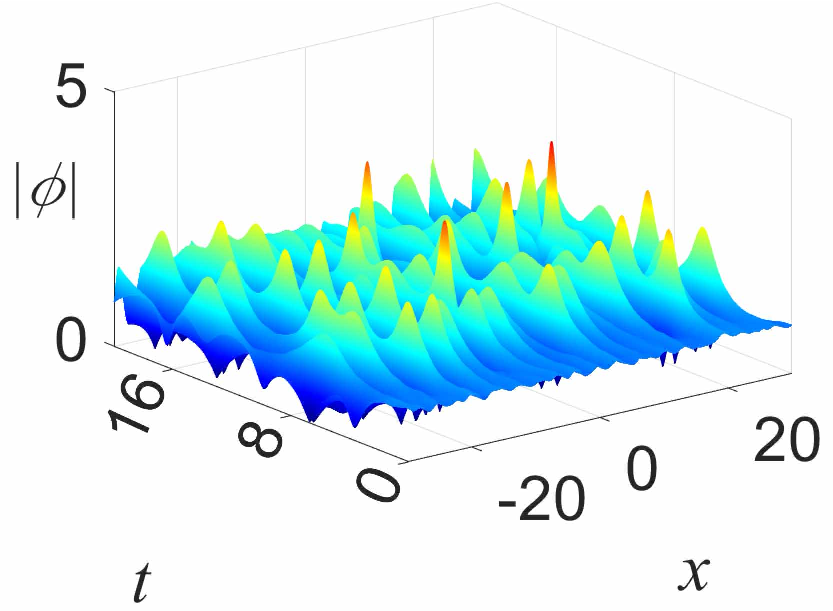} %
\includegraphics[height=95pt,width=105pt]{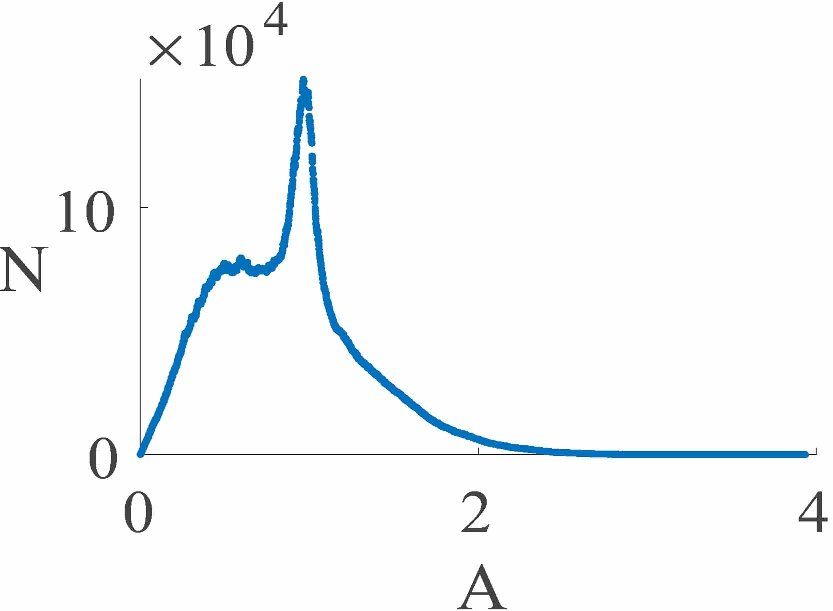} \newline
\caption{Dissipative RWs in the case of the focusing nonlinearity ($\protect%
\sigma =-1$). The left panel: RWs in the chaotic state exhibited by $|%
\protect\phi (x,t)|$, as produced by the simulations with random
perturbations added to the stationary \textquotedblleft dnoidal" state. The
parameters are $\Delta =-2$, $k=0.3$, $\protect\gamma =0.2$. The right
panel: the probability distribution of number $N$ of extreme events as a
function of the corresponding local amplitude $A$ of the chaotic field.}
\label{fig6}
\end{figure}

It is well known that MI can lead to the generation of RWs~\cite{nail1,rn2}.
Because the dnoidal waves are modulational unstable, we examine here if such
instability can trigger the formation of RWs. We start the respective
simulations, using the dnoidal states perturbed by small-amplitude white
noise:
\begin{equation}
\phi (x,0)=\phi _{\mathrm{dn}}(x)+\epsilon f(x),  \notag
\end{equation}%
where $\epsilon $ is a small real amplitude of the noise, and $f(x)$ is a
uniformly distributed complex function, whose real and imaginary parts have
random values in the interval of $[-1,+1]$. We used the split-step Fourier
algorithm, solving the linear and nonlinear parts of Eq. (\ref{lle}) by
means of the Fourier transform and Runge-Kutta method, respectively. Adding
the small random perturbations to the dnoidal input~(\ref{dns}) leads to a
chaotic state created by the asymmetric MI. The spectral component of the
noise with the highest growth rate dominates and splits the dnoidal wave
into localized modes, while the presence of other frequencies leads to
chaotic evolution of those modes. In this chaotic state, one can identify a
few events with the local intensity significantly larger than the average
value, as shown in the left panel of Fig.~\ref{fig6}.

The probability of having a wave with a certain peak amplitude in the
chaotic wave field provides the essential information about the formation of
the RWs. We studied these probabilities for the emerging chaotic wave field,
using the largest available volume of numerical data. We counted local
maxima of the chaotic field in a spatial domain of $[-60L,+60L]$, discarding
ones with the amplitude $<0.001$. The right panel of Fig.~\ref{fig6} shows
the probability distribution of the number of such events as a function of
the respective amplitude, which exhibits a long-tail probability
distribution. The respective SWH value is $1.3479$, with some events
featuring the amplitude exceeding the double SWH. Thus, the statistical
analyses show that the amplitude probability distribution possesses a long
tail with the intensity height well beyond the double SWH, clearly implying
that these extreme events belong to the class of RWs.

\section{The transverse instability and formation of MSEs (multi-spot
excitations) in the 2D regime}

Our objective here is to study the TI and subsequent evolution of unstable
quasi-1D patterns in the framework of the 2D LL equation, but still driven
by the 1D SPP:
\begin{equation}
i\left( \gamma +\frac{\partial }{\partial t}\right) \phi =\left[ -\frac{1}{2}%
\left( \frac{\partial ^{2}}{\partial x^{2}}+\frac{\partial ^{2}}{\partial
y^{2}}\right) +\Delta +\sigma |\phi |^{2}\right] \phi +E(x).  \label{lle2}
\end{equation}%
We consider the TI of the above-mentioned quasi-1D stationary states against
2D perturbations, adding them to the unperturbed state as
\begin{eqnarray}
&&\hspace{-0.4cm}\phi (x,y,t)=e^{i\theta }\psi (x)+\\
&&\epsilon \left[ (u(x,\rho )e^{\lambda
t+i\rho y}+\mathrm{c.c.})+i(v(x,\rho )e^{\lambda t+i\rho y}+\mathrm{c.c.})%
\right],\nonumber
\end{eqnarray}%
with a transverse wavenumber $\rho $, cf. its 1D counterpart, defined above
by Eq. (\ref{lsp0}). Substituting this expression in Eq.~(\ref{lle2}), one
arrives at the linear eigenvalue problem [cf. its 1D counterpart (\ref{lsp1}%
)]:%
\begin{equation}
\lambda \left[
\begin{array}{l}
u(x,\rho ) \\
v(x,\rho )%
\end{array}%
\right] =\left[
\begin{array}{rr}
\mathcal{M}_{11} & \mathcal{M}_{12} \\
\mathcal{M}_{21} & \mathcal{M}_{22}%
\end{array}%
\right] \left[
\begin{array}{l}
u(x,\rho ) \\
v(x,\rho )%
\end{array}%
\right] ,  \label{lsp2}
\end{equation}%
with matrix elements $\mathcal{M}_{11}=-\gamma +\sigma \sin (2\theta )\psi
^{2}$, $\mathcal{M}_{12}=\Delta -\sigma \cos (2\theta )\psi ^{2}+2\sigma
\psi ^{2}-(1/2)\partial _{x}^{2}+(1/2)\rho ^{2}$, $\mathcal{M}_{21}=-\Delta
-\sigma \cos (2\theta )\psi ^{2}-2\sigma \psi ^{2}+(1/2)\partial
_{x}^{2}-(1/2)\rho ^{2}$, and $\mathcal{M}_{22}=-\gamma -\sigma \sin
(2\theta )\psi ^{2}$.

\begin{figure}[tbp]
\centering
\includegraphics[height=95pt,width=105pt]{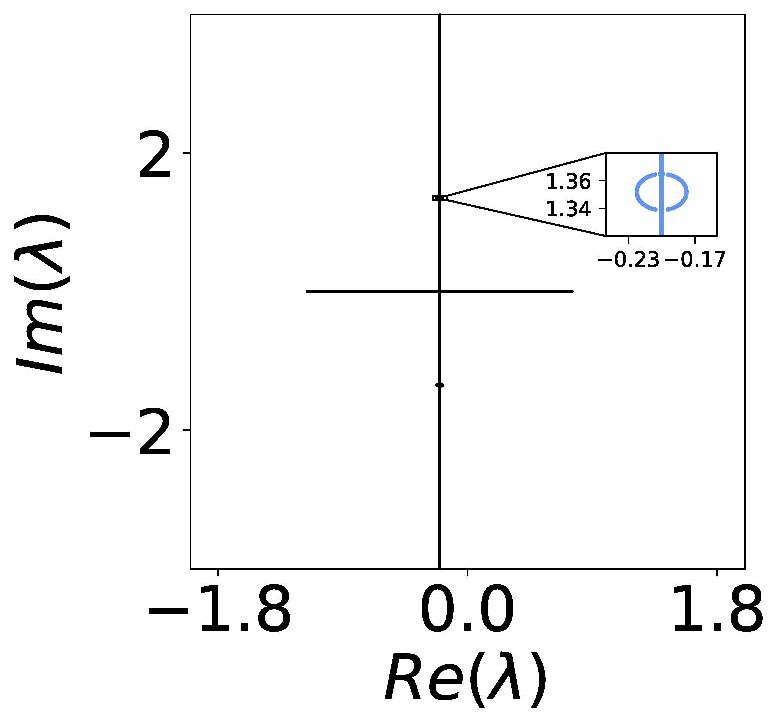}
\includegraphics[height=100pt,width=105pt]{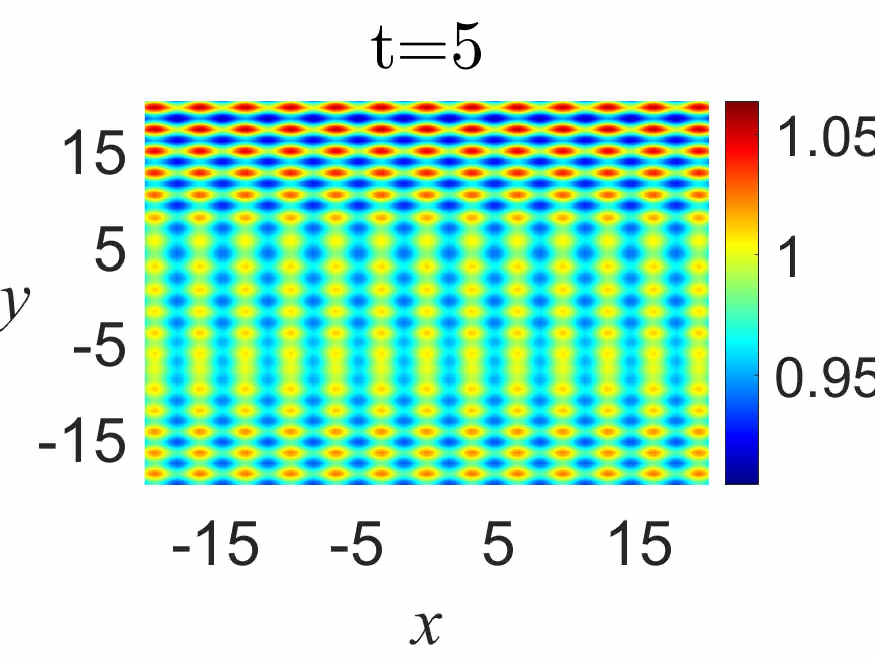} \newline
\caption{TI and the formation of 2D lumps. The left panel: the instability
spectrum produced by the numerical solution of Eq.~(\protect\ref{lsp2}) for
the underlying stationary state~(\protect\ref{dns}) with $\protect\sigma =-1$%
, $\Delta =-2$, $k=0.3$, $\protect\gamma =0.2$ and $\protect\rho =2$. The
right panel: an array of 2D lumps exhibited by the subsequent evolution of $|%
\protect\phi (x,t)|$, initiated by the stationary state~(\protect\ref{dns})
with $1\%$ random perturbations added to it.}
\label{fig7}
\end{figure}

Because of the similarity of the findings for the underlying cnoidal and
dnoidal states, we again focus on the TI analysis for the dnoidal solution~(%
\ref{dns}) in the case of the focusing nonlinearity, the left panel of Fig.~%
\ref{fig7} presenting at example. To study the nonlinear stage of the TI
development, we added initial perturbations at the $1\%$ level, which are
random with respect to both the longitudinal and transverse coordinates. It
has been found that, due to the TI, the dnoidal wave breaks into an array of
lumps, as shown in the right panel of Fig.~\ref{fig7}.

\section{Conclusion}

We have elaborated a scenario for the formation of MSEs (multi-spot
excitations) in Kerr cavities with SPP (spatially-periodic pump) in the
framework of the 1D and 2D LL (Lugiato-Lefever) equations. Looking for SPPs
in the form of Jacobi's elliptic functions, we have produced three exact 1D
solutions, in the form of snoidal or cnoidal/dnoidal stationary patterns, in
the case of the defocusing or focusing sign of the cubic nonlinearity,
respectively. In the 1D regime, the analysis has produced MI (modulational
instability) in the form which is asymmetric with respect to the imaginary
axis in the stability-eigenvalue plane. The MI is suppressed when the loss
constant in the LL equation exceeds a certain threshold value. In the case
of the defocusing nonlinearity, the development of the MI splits the snoidal
pattern into an array of spatiotemporal crescents, while in the case of
self-focusing, depending on the choice of initial perturbations, MI creates
dissipative RWs (rogue waves) or chains of breathers, whose evolution
features and FPUT (Fermi-Pasta-Ulam-Tsingou)-like recurrence. In the 2D
regime, the development of TI (transverse instability) of the exact
spatially periodic stationary solutions gives rise to arrays of 2D lumps.
This study provides new insights into the complex MSE dynamics in Kerr
cavities under the action of the interplay of SPP and losses. The findings
reported here provide a contribution to nonequilibrium physics in general,
and they may find\ direct applications in laser physics.

A relevant subject for a follow-up work is to construct numerical spatially
periodic solutions, in which all basic parameters (in particular, the
spatial period, determined by modulus $k$ in the analytical solutions) may
be varied independently, to facilitate comprehensive studies of the
dynamical behavior produced by the LL equation. Another challenging topic is
to study the stability problem for the 2D LL equation with a 2D structure of
the spatially periodic pumps.\vspace{2mm}

\textit{Acknowledgments.} This work has been supported by the Fundamental
Research Funds of the Central Universities (No. 230201606500048) and Israel
Science Foundation (grant No. 1695/22).

\nocite{}

\end{document}